\documentclass[12pt,thmsb,titlepage]{article}
\usepackage{amsmath}
\usepackage{amssymb}
\usepackage{graphicx}
\usepackage{epsf}
\usepackage[T1]{fontenc}
\usepackage[latin9]{inputenc}
\usepackage[letterpaper]{geometry}
\usepackage{setspace}
\usepackage{epstopdf}

\usepackage{float}
\usepackage{rotating}

\usepackage{epstopdf}

\providecommand{\tabularnewline}{\\}

\usepackage{epsf}
\usepackage[hidelinks]{hyperref}
\usepackage{amsmath}
\usepackage{amssymb}
\usepackage{graphicx}
\newcommand{\E}{\mathbb{E}}

\usepackage{caption}
\usepackage{subcaption}

\newcommand{\captionfonts}{\footnotesize}
\makeatletter  
\long\def\@makecaption#1#2{%
  \vskip\abovecaptionskip
  \sbox\@tempboxa{{\captionfonts #1: #2}}%
  \ifdim \wd\@tempboxa >\hsize
    {\captionfonts #1: #2\par}
  \else
    \hbox to\hsize{\hfil\box\@tempboxa\hfil}%
  \fi
  \vskip\belowcaptionskip}
\makeatother   

\newtheorem{theorem}{Theorem}

\newtheorem{lemma}{Lemma}

\newtheorem{assumption}{Assumption}
\usepackage{color}

\usepackage[authoryear]{natbib}
\begin{document}

\centerline{\Large{\bf Weak Identification with Many Instruments }}

\medskip

\centerline{Anna Mikusheva\footnote{
Department of Economics, M.I.T. Address: 77 Massachusetts Avenue, E52-526, Cambridge, MA, 02139. Email: amikushe@mit.edu. } and Liyang Sun\footnote{UCL and CEMFI. Email: lsun20@cemfi.es Liyang Sun gratefully acknowledges support from Ayudas Juan de la Cierva Formaci\'on. The paper benefited from
the comments of conference and seminar audiences at the RES 2023, Stone Centre at UCL, Boston College, University of Arizona, SEA 2023, University of Bristol, and UC3M.  We are grateful to Mikkel S{\o}lvsten and Tiemen Woutersen for advice,  to John C. Chao and Brigham Frandsen for sharing code for simulations. The accompanying Stata package \texttt{manyweakiv} is available at \url{https://github.com/lsun20/manyweakiv} }} 
\medskip

\centerline{\bf Abstract}
\smallskip

\noindent {\small{Linear instrumental variable regressions are widely used to estimate
causal effects. Many instruments  arise from the use of ``technical'' instruments and more recently from the empirical strategy of ``judge design''.  This paper surveys and summarizes ideas from recent literature on estimation and statistical inferences with many instruments for a single endogenous regressor. We discuss how to assess the strength of the instruments and how to conduct  weak identification-robust inference  under heteroskedasticity. We establish new results for a jack-knifed version of the Lagrange Multiplier (LM)  test statistic.  Furthermore, we  extend the weak-identification-robust tests to settings with both many exogenous regressors and many instruments. We propose a test that  properly partials out many exogenous regressors while preserving the re-centering property of the jack-knife.  The proposed tests have  correct size and good power properties. 
}}
\smallskip

 \noindent
     \textsc{Keywords:} instrumental variable regressions, many instruments, weak instruments
     
     \bigskip
\section{Introduction}

In linear instrumental variables (IV) regression, when there are many 
instruments, the consistency of the estimation for the first stage coefficients becomes questionable. If the uncertainty
about the first stage coefficients has a first order importance, conventional approximations to the distribution of IV estimators are generally unreliable.  Recognizing this problem, \cite{bekker_alternative_1994} formally modeled the issue of many instruments by considering asymptotic approximations that assume the number of instruments grows to infinity with the sample size. Specifically, \cite{bekker_alternative_1994} is  the first paper that pointed out  the standard two-stage least squares (TSLS) estimator can be badly biased under many instruments.

Our paper provides an exposition of the challenges discovered and some solutions  proposed  in the three decades since the original paper of  \cite{bekker_alternative_1994} in fast-growing econometric literature on estimation and statistical
inferences (testing, confidence sets construction) in linear IV models with many 
instruments and a single endogenous regressor. We first describe the trade-offs that arise from using many instruments. The benefit of using more instruments is obvious --- they bring additional exogenous information that can help to estimate the structural parameter of interest and may lead to a more efficient estimator. The challenges of using many instruments arise from the need to find an optimal way to combine them (the task done by the first stage) and from the growing complexity of such a task. When the information from additional instruments grows slower than the complexity of the first stage, the additional instruments might be detrimental and lead to a worse estimator. Specifically, the uncertainty from the first stage tends to translate into the bias of the estimator for a structural parameter, and may lead to an inconsistency of the structural estimator.

We survey some influential ideas  attempting to properly use   information  from an increasing number of instruments. These ideas include jack-knifing and sample splitting; they produce  estimators  with a superior performance in comparison to the TSLS estimator in settings with many instruments. We then discuss the definition of weak identification, a situation when the information contained in the instruments is low relative to the number of instruments to the extent that conventional approximations to the distribution of IV estimators become invalid. We describe an empirically relevant  pre-test for weak identification as well as identification robust tests.

In addition to a survey of the existing literature,  we also establish two new results related to identification robust tests. First of all, we present a new form of identification robust LM test that uses a new estimator of variance. Our test has superior power properties in comparison with the LM test recently proposed in \cite{matsushita_otsu_2022}. The second new result is an identification robust Anderson-Rubin (AR) test that is valid under both  many instruments and many exogenous regressors (controls).

Throughout this paper, we maintain the assumption of homogeneous treatment effect. However, under a proper monotonicity assumption, results from  \cite{kolesa2013estimation} show that two-step estimators are all consistent for a convex combination of local average treatment effects. Therefore, some results presented in this paper can potentially be extended to allow for heterogeneous treatment effects.

The remainder of this paper is organized as follows. Section 2 summarizes results on estimation with many instruments and defines weak identification. Section 3 describes a pre-test for weak identification. Section 4 describes existing weak identification robust tests and introduces the new LM test. Section 5 discusses the challenges and solutions of having many exogenous regressors. Section 6 concludes with open questions.

We finish the introduction by presenting two well-known examples of many instruments. These examples demonstrate that whenever there is a good exogenous variation allowing the identification of causal effects, many instruments arise naturally. 

\textbf{Example 1}: \cite{AngristKrueger1991} contains one of the most well-known applications of linear IV regression. This paper estimated the return to education using the quarters of birth as instruments, and was prominently mentioned in the press-release about the  Sveriges Riksbank Prize in Economic Sciences in Memory of Alfred Nobel 2021. The structural parameter of interest is  the  coefficient on the educational attainment in the structural equation:
$$            wage_i=\beta~education_i+controls_i+e_i.  $$
Due to compulsory educational laws, children are required  to stay in school until a certain physical age, and yet children typically start their schooling in September when they are at a different physical age. Thus, the quarter of birth, which can be considered as randomly assigned in the population, produces an exogenous and observable variation in  educational attainment. Since the compulsory educational laws vary by  state, the effects of the quarter of birth on education are heterogeneous across the states. The effects are possibly heterogeneous by birth cohort as well. Therefore, \cite{AngristKrueger1991} considered specifications that interact quarter of birth dummies   with either state or year of birth dummies or both. In the main specifications used by  \cite{AngristKrueger1991}, interactions of the quarter of birth  with the year of birth dummies yield 30 instruments. Adding interactions with state of birth dummies yields 180 instruments. Finally, adding fully saturated three-way interactions yields 1530 instruments. 

In this setting, one starts with a single variable producing the exogenous variation, but creates multiple ``technical'' instruments based on it in order to extract information from heterogeneous and/or non-linear first stage. Such a setting is extremely common in empirical practice and is one of the main sources of examples with many instruments. Not only did \cite{AngristKrueger1991} inspire the  literature on weak identification, as it was the main example in \cite{StaigerStock1997}, it was also a motivating example for the literature on  many instruments including \cite{hansen_estimation_2008}.

\textbf{Example 2.} ``Judge design'' is a common name for empirical settings that use the exogenous assignment of cases  to different decision makers as instruments for a treatment in an attempt to estimate important causal effects of interest. Fueled by rich administrative data, recent applications of ``judge design'' include \citet{Maestas_Does_2013,Dobbie_Effects_2018,Sampat_How_2019}, and \cite{Bhuller_Incarceration_2020}. For example, \cite{Bhuller_Incarceration_2020} estimate the effect of incarceration on  recidivism using random assignments of criminal cases to judges as a source of exogenous variation. Judges express different leniencies  producing an exogenous variation in incarceration decisions. The instruments here are dummies for individual judges. Since each judge can only process a certain number of cases out of the total court cases, the number of judges (the number of instruments) increases fast with the sample size. Similar situation arises in applications of Mendelian randomization \citep{davies2015} and   name-based estimators of inter-generational mobility  \citep{torsten2024}, among others.

\section{Estimation with many instruments}

To describe the main ideas, we consider a simplified setup with one endogenous regressor and no included exogenous regressors (controls). We will add exogenous regressors in Section \ref{section- covariates}. Assume we observe an i.i.d.  sample $\{(X_i,Y_i,Z_i), i=1,...,N\}$ satisfying a linear IV model: 
$$
 \left\{ \begin{array}{c}
           Y_i=\beta X_i+e_i; \\
           X_i=\pi'Z_i+v_i,
         \end{array}
 \right.
 $$
 where $X_i$ is a one-dimensional endogenous regressor, $Z_i\in\mathbb{R}^K$ are instruments satisfying the exclusion restriction $\E[e_i|Z_i]=0$. Additionally, we assume a linear first stage, $\E[v_i | Z_i] = 0$, which simplifies exposition and is crucial for deriving certain asymptotic distributions -  particularly, the LM statistics discussed later heavily rely on this assumption. Notably, this assumption is not required for the AR test introduced later. We allow errors to be heteroskedastic with $0<c<\E[e_i^2|Z_i]<C$. In this setting we are interested in estimation of and statistical inferences on the structural coefficient $\beta$, while $\pi$ is a set of nuisance parameters. The first equation is often referred to as the structural equation, while the second is called the first stage. We will discuss estimation and statistical inference conditional on the realization of $Z_i$, and thus will treat instruments as fixed. For  simplicity of notation, we  drop the conditioning sign and all expectation signs should be read as conditional on $Z_i$'s.

 \subsection{Asymptotic bias of TSLS}\label{section - bias of TSLS}
 
The TSLS is the most widely known and used estimator in this case. To implement the TSLS, one first runs the OLS regression of the first stage by regressing $X_i$ on $Z_i$ which obtains the estimated coefficients $\widehat\pi$. Then one runs the second stage regression of $Y_i$ on $\widehat X_i=\widehat\pi'Z_i$, where the regression coefficient estimate $\widehat\beta_{TSLS}$ is the TSLS estimate. 

The TSLS estimator is an efficiently weighted GMM estimator under conditional homoskedasticity of errors $e_i$ and thus possesses the asymptotic optimality under homoskedasticity in a setting with a small number of instruments.  The notion of asymptotic efficiency for GMM appeared in \cite{chamberlain_1987}.  Since the unknown parameter $\beta$ is a scalar, a single relevant instrument is sufficient for identification. If we have $K$ instruments and can use any linear combination of them as the single instrument, the question is which linear combination provides an estimator with the smallest asymptotic variance. In the homoskedastic model the optimal combination is $\E[X_i|Z_i]=\pi'Z_i$ as  it delivers  asymptotic efficiency.  An alternative interpretation of the TSLS estimator through the lens of ``optimal instruments'', is that the TSLS estimator uses the first stage to combine multiple instruments $Z_i$ into a single estimated ``optimal'' instrument $\widehat X_i$. One can show that $\widehat\beta_{TSLS}$ is equal to the IV estimate in a just-identified IV regression of $Y_i$ on $X_i$ using $\widehat X_i$ as the single instrument. Furthermore, the asymptotic variance of the TSLS estimator as well as of the infeasible optimal IV estimator is inversely proportional to $\pi'Z'Z\pi$, the  part of the endogenous regressor $X$ explained by $Z$. Thus, if we know how to construct the optimal instrument from the available data, we can expect additional instruments to improve efficiency of the TSLS estimator through increasing the explained part of the endogenous regressor.

In practice we do not know $\pi$, the coefficient of the optimal instrument combination, and have to estimate it in the first stage. As we show, this typically leads to a bias in the estimation of the structural coefficient that increases with the number of instruments.  Specifically, the estimated optimal instrument 
$$
\widehat{X}_i=X'Z(Z'Z)^{-1}Z_i=\pi'Z_i+v'Z(Z'Z)^{-1}Z_i
$$
contains not only the true optimal instrument $\pi'Z_i$ but also the estimation mistake, which makes the estimated optimal instrument endogenous. For the next derivation only,  assume that the errors $e_i$ and $v_i$ are homoskedastic with $\sigma_{ev}=\E[e_iv_i]\neq 0$. The parameter $\sigma_{ev}$  measures the degree of endogeneity of the regressor, and therefore the bias of the OLS. Then
$$
\mathbb{E}\left[\frac{1}{N}\sum_{i=1}^{N}(\widehat{X}_i-\pi'Z_i)e_i\right]=\frac{1}{N}\sum_{i=1}^{N}Z_i'(Z'Z)^{-1}Z_i\mathbb{E}[v_ie_i]= \frac{tr(Z(Z'Z)^{-1}Z')}{N}\sigma_{ev}=\frac{K}{N}\sigma_{ev}.
$$
As shown above, the estimated instrument $\widehat{X}_i$ is endogenous, and consequently the TSLS has bias proportional to a $\frac{K}{N}$. Specifically, the bias of TSLS is increasing in the number of instruments $K$, and if the number of instruments is a large fraction of the sample size, then the TSLS is inconsistent. This result first appeared in \cite{bekker_alternative_1994}, who argues that in order to get a more realistic approximation for the properties of the TSLS estimator when the number of instruments is large, it is important to model the number of instruments as increasing with the sample size.

Another explanation for the TSLS bias is over-fitting of the first stage when the number of  regressors in the first stage regression is large. Suppose there are $N$ instruments for a sample of size $N$, then the first stage regression would produce a perfect fit and we have $\widehat X_i=X_i$. In this case, the TSLS equals to the OLS, which is inconsistent and biased due to endogeneity of $X$. This   is an extreme case, but it provides an intuition for why having many instruments may complicate estimation.

In order to avoid the over-fitting problem, one may consider using some alternative estimation strategies for the first stage.  \cite{donald_choosing_2001} proposed an instrument selection procedure based on a Mallows criteria. Suggestions to use LASSO selection on the first stage were put forward by \cite{belloni_sparse_2012}  and 
\cite{belloni_inference_2014}. \cite{okui_instrumental_2011} proposed to use a shrinkage estimator, while \cite{carrasco_regularization_2012} suggested several regularization procedures based on the
spectral decomposition of the conditional expectation operator, such as the principal components approach and Tikhonov's regularization. 

If one is willing to impose some assumptions about the form
of the optimal instrument, then with a proper estimation technique on the first stage that allows for a  consistent estimation of the optimal instrument, one may obtain a semi-parametric efficient estimator for $\beta$. For example, \cite{donald_choosing_2001} assumed a
known ordering among instruments (or groups of instruments) by strength/informativeness. The
LASSO procedure of \cite{belloni_sparse_2012} delivers  a semi-parametric efficient estimator for $\beta$  if the first-stage
regression is approximately sparse, that is, a relatively small number of the instruments
successfully approximates the optimal instrument. 
Another type of assumption often needed is a regularity condition placed on the conditional expectation operator. For example, \cite{belloni_sparse_2012} restricted eigenvalues of
an empirical Gram matrix, while \cite{carrasco_regularization_2012} assumed that the conditional expectation operator is a Hilbert-Schmidt operator. 

When the assumptions about the form of the optimal instrument fail, the performance of these alternative estimation strategies are not always guaranteed. \cite{HansenKozbur2014} provided simulation evidence that the performance of IV estimators using LASSO in the first stage  is less than stellar when the signal
on the first stage is dense and weak. \cite{angrist_machine_2022}  studied the performance of some Machine Learning (ML) techniques for instrument selection using simulations calibrated to two important empirical examples. They compared the performance of
IV estimators using LASSO and random forest in the first stage with that of OLS, TSLS and several jack-knife and split-sample estimators that we will
discuss below. In almost all cases the IV estimators using LASSO and random forest
in the first stage resulted in biases whose magnitudes are comparable to those of OLS and TSLS without much
improvement in variance. Moreover, the performance of these two ML methods depends heavily on the choice of the regularization parameter: the cross-validation
or plug-in penalties for the LASSO, or the leaf-size for the random forest. None of
the standard choices for the regularization parameter were totally satisfactory. One plausible explanation is that the sparsity of the first stage is a poor description of the data in these two empirical applications. 

\subsection{Jack-knifing or diagonal removing}
The bias of the TSLS arises from using the same observation in both stages of estimation. Since the first stage estimate $\widehat\pi$ depends on $X_i$, the estimated optimal instrument $\widehat X_i$ used by the TSLS is  
 endogenous and  is correlated with the structural error $e_i$. In this subsection, we survey some influential ideas attempting to remove the bias by jack-knifing and sample-splitting. 
 
 \cite{angrist_split_1995}  proposed removing the bias by using separate samples in the two stages of the TSLS. They suggested splitting the original sample into two halves. If $\pi$ is estimated  using the first half of the sample, and the estimated optimal instrument is produced for the second half, then the constructed instrument will be exogenous and the IV estimator would avoid the over-fitting bias of the TSLS. This idea is called sample-splitting.

A refinement of sample-splitting that exploits the data in a more sophisticated way is jack-knifing  \citep{angrist_jackknife_1999}. The idea is to run a separate first stage for each observation. Namely, for observation $i$ we run the OLS regression of $X$ on $Z$ on the sample excluding observation $i$, calling the resulting estimate $\widehat\pi_{(-i)}$. Define $Z_i^*=\widehat\pi_{(-i)}'Z_i$ and run the OLS regression of $Y_i$ on $X_i$ using $Z_i^*$ as the single instrument. The resulting estimate of $\beta$ is called a jack-knife IV estimator (JIVE). JIVE breaks the dependency between two stages on the same observation and effectively satisfies the exogeneity condition $\E[Z_i^*e_i]=0$ for the constructed instrument.

The idea of running a separate first stage OLS regression for each observation seems daunting but is ultimately unnecessary. There is a relatively easier  formula for JIVE based on the Sherman-Morrison-Woodbury formula, which provides an explicit way to calculate the leave-one-out projection based on the full-sample orthogonal projection. Specifically, one can show that while the TSLS can be written as
$$
\widehat\beta_{TSLS}=\frac{X'P_ZY}{X'P_ZX}=\frac{\sum_{i,j}P_{ij}X_{i}Y_j}{\sum_{i,j}P_{ij}X_{i}X_j},
$$
where $P_Z=Z(Z'Z)^{-1}Z'$ is a projection on $Z$, and $P_{ij}$ are its elements, the JIVE has a similar form:
\begin{equation}
\widehat\beta=\frac{\sum_{i,j}P^*_{ij}X_{i}Y_j}{\sum_{i,j}P^*_{ij}X_{i}X_j},\label{eq:jiv1}
\end{equation}
where elements $P_{ij}^*$ are slightly re-weighted elements of $P_{ij}$ with one important difference - all diagonal elements are zeros. 

The diagonal elements of the projection matrix are tightly connected to the TSLS bias, since the expectation of the  numerator for  $\widehat\beta_{TSLS}-\beta$ is
$$
\E\left[\sum_{i,j}P_{ij}X_{i}e_j\right]=\E\left[\sum_{i,j}P_{ij}v_{i}e_j\right]=\sum_{i}P_{ii}\E\left[v_{i}e_i\right].
$$
Therefore, a closely related estimator, which just removes the diagonal from the TSLS formula, is also often referred to as JIVE:
\begin{equation}
 \widehat\beta_{JIVE}=\frac{\sum_{i\neq j}P_{ij}X_{i}Y_j}{\sum_{i\neq j}P_{ij}X_{i}X_j}. \label{eq:jiv2}  
\end{equation}
This estimator was proposed in \cite{angrist_jackknife_1999} and called JIV2 by the authors. It is numerically extremely close to the other JIVE described in \eqref{eq:jiv1}. For simplicity we will call  the estimator $\widehat\beta_{JIVE}$ defined in \eqref{eq:jiv2} the JIVE. The expectation of the numerator of the estimation error for the JIVE,
$$
\widehat\beta_{JIVE}-\beta=\frac{\sum_{i\neq j}P_{ij}X_{i}e_j}{\sum_{i\neq j}P_{ij}X_{i}X_j},
$$
is zero, $\E\left[\sum_{i\neq j}P_{ij}X_{i}e_j\right]=0$, and thus, the JIVE avoids the over-fitting bias of the TSLS estimator.

The JIVE can also be motivated as the optimizer of a slightly corrected objective function, specifically:
$$
\widehat\beta_{JIVE}=\arg\min_\beta Q_{JIVE}(\beta)=\arg\min_\beta\sum_{i\neq j}P_{ij}(Y_i-\beta X_i)(Y_j-\beta X_j).
$$
The JIVE objective function is only slightly different than the TSLS objective function:
$$
Q_{TSLS}(\beta)=\sum_{i, j}P_{ij}(Y_i-\beta X_i)(Y_j-\beta X_j)=(Y-\beta X)'P_Z(Y-\beta X).
$$
As was pointed out in \cite{han_gmm_2006} and \cite{NeweyWindmeijer2009},  the problem with the TSLS objective function is that its expectation at the true parameter value is not zero, $\E Q_{TSLS}(\beta_0)\neq 0$, and therefore is not minimized at the true parameter value. The JIVE objective function solves this issue by removing the diagonal. This approach can be extended to other instrumental variable estimators for bias reduction. For instance, \cite{hausman_instrumental_2012} proposed a version of the LIML and Fuller estimators with the diagonal removed (JIVE-LIML and JIVE-Fuller), while \cite{HansenKozbur2014} proposed a similar modification for  a ridge estimator.

\subsection{Consistency of estimators with many instruments}

JIVE-type estimators have superior consistency properties when compared with the TSLS. In particular, \cite{chao_consistent_2005} established that under homoskedasticity the TSLS is consistent when $\frac{\pi'Z'Z\pi}{K}\to\infty$, while the JIVE is consistent when $\frac{\pi'Z'Z\pi}{\sqrt{K}}\to\infty$. These are drastically different conditions  when the number  of instruments $K$ is growing with the sample size. A similar statement under heteroskedasticity appeared in \cite{hausman_instrumental_2012} along with consistency of JIVE-LIML and JIVE-Fuller when $\frac{\pi'Z'Z\pi}{\sqrt{K}}\to\infty$. 

Notice that $\pi'Z'Z\pi$, the explained part of regressor $X$,  measures the information contained in the optimal instrument. If one knew the optimal weights $\pi$, and used the TSLS with the single optimal instrument $Z\pi$, then this estimator is consistent as long as $\pi'Z'Z\pi\to\infty$. The TSLS bias is proportional to $K$ and constitutes the leading term causing inconsistency of the TSLS  when $\frac{\pi'Z'Z\pi}{K}$ is asymptotically bounded. Once the bias is removed, the consistency arises when the next asymptotic term is negligible.

The condition for consistency of JIVE ($\frac{\pi'Z'Z\pi}{\sqrt{K}}\to\infty$) emphasizes that additional instruments do not just increase the information extracted from the sample ($\pi'Z'Z\pi$) but also come with a cost. To justify adding new instruments, they should bring enough additional information so that $\frac{\pi'Z'Z\pi}{\sqrt{K}}$ increases. This happens since the optimal coefficients $\pi$ are not known and have to be estimated. The estimation of the optimal coefficients always comes with mistakes, which accumulate with the dimensionality of instruments, $K$. The factor $\sqrt{K}$ is the price one pays for the need to search for an optimal combination in a  high-dimensional setting.

In \cite{mikusheva_inference_2022} we showed that this price is unavoidable if  the direction or form of the optimal instrument is fully unknown. Specifically, we showed that if coefficients $\pi$ are completely unknown and $\frac{\pi'Z'Z\pi}{\sqrt{K}}$ is asymptotically bounded, then for any two distinct values of parameter $\beta$ there exists no asymptotically consistent test distinguishing them. This implies that $\frac{\pi'Z'Z\pi}{\sqrt{K}}\to\infty$ is a necessary condition for consistency.

However, if one is willing to impose assumptions on the form of the optimal instrument, then the  condition for consistency may be weakened. Specifically, if one is willing to assume that the optimal combination is sparse and uses a properly chosen LASSO procedure  on the first stage, then the condition for consistency may be weakened to depend (up to logarithm multipliers) on the squared root of the sparsity parameter in place of $\sqrt{K}.$

\cite{chao_asymptotic_2012}  and \cite{hausman_instrumental_2012} established that under some minor assumptions, mainly additional moment restrictions and assumptions that the projection operator $P_Z$ is well-balanced, the condition $\frac{\pi'Z'Z\pi}{\sqrt{K}}\to\infty$ implies that the JIVE, JIVE-LIML and JIVE-Fuller are asymptotically Gaussian.  This implies that one can employ t-statistics and produce Wald-type confidence sets, so the inferences are somewhat standard in this case. The one caveat of this statement is that the usual formulas for standard errors are incorrect and understate the uncertainty. These papers also proposed the consistent asymptotic variance estimators.

\section{Pre-testing for  identification strength}\label{section: pre-test}

The assumption that the amount of information extracted by the instruments  is sufficiently large  relative to the squared root of the number of instruments  ($\frac{\pi'Z'Z\pi}{\sqrt{K}}\to\infty$), 
 along with some technical conditions,  guarantees the consistency of the JIVE. Under the same assumption a properly defined $t$-statistic based on the JIVE converges asymptotically to a standard Gaussian distribution, making the usual $t$-test valid.  However, if this assumption does not hold, then the JIVE and other JIVE-type estimators tend to be biased, and the asymptotic distribution of their  $t$-statistics cannot be reliably approximated by a standard Gaussian distribution. Consequently,  the $t$-tests and  confidence sets based on the $t$-statistics are asymptotically  invalid. This  phenomenon is known as weak identification, as explained in \cite{mikusheva_inference_2022}.

Weak identification has been recognized in IV estimation with small number of instruments by \cite{StaigerStock1997}  and \cite{StockYogo2005}. Stock and Yogo proposed a pre-test for weak identification, which compares the first stage $F$-statistics for the hypothesis that $\pi=0$ with a specially selected cut-off. The cut-off technically depends on the number of instruments and the goal of the pre-test (to bound the bias of the TSLS estimator or to bound the size distortions of the $t$-test), but in practice the universal and simple cut-off of 10 was suggested and has been widely adopted. In empirical research, whenever the first stage $F$ exceeds 10,  it is  considered reliable to use the standard TSLS inferences, while otherwise one should employ an identification robust inference.

As pointed out by \cite{hansen_estimation_2008}, this pre-test does not work well in the case of many instruments. Specifically, the first stage $F$ pre-test seems to indicate weak identification in a wide range of cases where a reliable estimator exists. This is mainly because the first stage $F$ pre-test was built to assess the quality of inference procedures based on the TSLS estimator. As was shown in the previous section, the TSLS estimator is a poor choice of estimator for a case with many instruments. Figure 4 in \cite{StockWrightYogo2002} explicitly showed that the proper cut-off for the first stage $F$ statistics should depend in a significant way on the estimator used when the number of instruments is larger than 5 and should decline fast with the number of instruments for the JIVE and some other bias-corrected estimators. 

In \cite{mikusheva_inference_2022}, we proposed a new  pre-test for weak identification that is aimed at assessing the validity  of the JIVE $t$-statistic inferences in cases of many instruments, while allowing for a general form of heteroskedasticity. 

The logic behind such a pre-test is very similar to that behind the first stage $F$ pre-test in settings with a small number of instruments. First, we derived the distribution of the JIVE $t$-statistics under some technical conditions but without imposing the assumption that $\frac{\pi'Z'Z\pi}{\sqrt{K}}\to\infty$. We then isolated the parameter that directly quantifies the deviations of the JIVE $t$-statistics' asymptotic distribution from the standard Gaussian. This theoretical parameter is an analog of the concentration parameter in the IV setting with a small number of instruments introduced by \cite{StaigerStock1997}. We showed that this parameter is $\frac{\mu^2}{\Upsilon\sqrt{K}}$, where $\mu^2=\sum_{i\neq j}P_{ij}(\pi' Z_i)(\pi'Z_j) $ is a diagonal-removed version of $\pi'Z'Z\pi$, while $\Upsilon$ is a measure of the first stage uncertainty analogous to the variance of the first stage error in the homoskedastic case. Notice that this parameter has $\sqrt{K}$ in the denominator, as we would expect for the case of many instruments.
Then depending on how much  distortion from the declared size a researcher agreed to tolerate, we derive a cut-off for the theoretical parameter.  For example, for a 5\%-test and a tolerance for 5\% distortion, the cut-off is 2.5, which means that if $\frac{\mu^2}{\Upsilon\sqrt{K}}>2.5$, then the JIVE $t$-test of the nominal 5\% size cannot have an asymptotic size of above 10\%. 
Finally, we proposed an estimator $\widetilde F$ for the theoretical parameter $\frac{\mu^2}{\Upsilon\sqrt{K}}$ and derived its accuracy to create the cut-off for statistics $\widetilde F$. For example, if $\widetilde{F}>4.14$, then  with 95\% confidence $\frac{\mu^2}{\Upsilon\sqrt{K}}>2.5$, and the JIVE t-test has less than 5 \% size distortion. 

This pre-test for weak identification provides the following empirical guidance in settings with many instruments: use the JIVE $t$-test when  $\widetilde{F}>4.14$,  otherwise 
 use any of the weak identification robust tests discussed in the next section. This  two-step procedure mirrors the  approach popularized by \cite{StockYogo2005}. By Bonferroni inequalities, such two-step procedure guarantees that the total size of the two-step procedure is within 15\%. A more accurate analysis of the limit distribution in \cite{mikusheva_inference_2022} suggests that the maximal asymptotic size is bounded by 7\%.

An alternative to pre-test is to always report a robust confidence set. The next section discusses available approaches to constructing robust confidence sets. The Stata package \texttt{manyweakiv} implements the $\widetilde{F}$ pre-test for weak identification and several identification-robust tests detailed here. Additional information on the package is available in \cite{Sun_manyweakiv_2023}.

\paragraph{Empirical example: \cite{AngristKrueger1991}.} In  the analysis of \cite{AngristKrueger1991}, \cite{StaigerStock1997} pointed out that the first stage $F$ statistic is low in the specification with many instruments   and suspected weak identification. \cite{hansen_estimation_2008}  argued that this is not the case of weak but rather many instruments.

\cite{mikusheva_inference_2022} formally assessed this question based on the $\widetilde{F}$ pre-test for weak identification created for a many instrument setting. We used the original data from \cite{AngristKrueger1991}, with a sample from the 1980 US census containing 329,509 men born 1930-39. We reproduced two specifications considered by \cite{AngristKrueger1991}. The first specification uses  180 instruments that include 30 interactions between quarter and year of birth dummies and 150 interactions between quarter and state of birth dummies. The second specification uses 1,530 instruments, the full set of three-way interactions among quarter of birth, year of birth and state of birth dummies. Table \ref{table: AK pretest} reproduces part of Table 1 of \cite{mikusheva_inference_2022}, reporting the first stage $F$, the $\widetilde{F}$ pre-test, the TSLS estimator, the JIVE,  and the Wald confidence set based on JIVE.  In calculation of all statistics and estimators, we partial out the exogenous regressors and then proceed as in a setting without exogenous regressors. This provides valid inferences given that this specification has a relatively small number (71) of exogenous regressors. In Section \ref{section- covariates} we discuss alternative approaches that should be taken when the number of exogenous regressors is large.

\begin{table}
\begin{centering}
\small{
\begin{tabular}{ccccccc}
\hline
 &FF &$\widetilde{F}$ &TSLS&JIVE& (std. error) & JIVE-Wald\tabularnewline
\hline
\hline
180 instruments & 2.4 &13.4 &0.083& 0.099&  (0.017)& {[}0.066,0.132{]} \tabularnewline
\hline
1530 instruments  & 1.3 & 6.2 & 0.063 & 0.072 & (0.025)&{[}0.024,0.121{]} \tabularnewline
\hline
\end{tabular}
}
\par\end{centering}

\caption{Pre-test results in an example based on \cite{AngristKrueger1991} }\label{table: AK pretest}

{\footnotesize{\emph{Notes: }Results on the first-stage $F$ statistics (FF), the pre-tests for weak identification, the JIVE and its standard error and the JIVE-Wald confidence sets for IV  specification underlying Table VII
Column (6) of \cite{AngristKrueger1991}. Sample size is 329,509. }}

\end{table} 

Based on the $\widetilde{F}$ pre-test for weak identification, the JIVE $t$-tests are reliable in both specifications with 180 and 1,530 instruments. At the same time the value of the first stage $F$ is low and suggests that  the TSLS and the TSLS-based confidence sets should not be trusted. Another important observation is that having many uninformative instruments may be detrimental to the statistical accuracy, as we see the specification with 1,530 instruments produced wider confidence sets.

\section{Identification robust tests}\label{section- robust tests}
When the pre-test for  identification strength indicates that the JIVE-inferences are not reliable, an important question is what valid statistical inferences can still be done. Low values of $\widetilde F$ suggest that information in the sample is low for the number of instruments used to the extent that the JIVE is inconsistent. This also implies that no other consistent estimator exists unless any additional information about the optimal instrument is available. However, even in the absence of a consistent estimator, we may still construct informative tests  and confidence sets that are asymptotically valid, in the sense that their probability of incorrectly rejecting the null hypothesis and covering the true parameter value, respectively, remains well-controlled. A statistical inference procedure that remains valid no matter  the identification strength is therefore called robust to weak identification.  A large literature has developed a variety of statistical inference procedures robust to weak identification when the number of instruments  is small. According to \cite{andrews_testing_2007} the same type of tests remain valid when the number of instruments slowly increases with the sample size ($K^3/N\to 0$). Here we discuss how one can refine the weak identification robust procedures to be robust under a large number of instruments, when the number of instruments may be a fraction of the sample size. 

\paragraph{Confidence sets.} While we focus on robust tests, if one is interested in creating a robust confidence set for $\beta$, this can be done by numerically inverting any of the robust  tests we discuss below. Specifically, one may conduct tests $H_0:\beta=\beta_0$ for different values of $\beta_0$ and collect the values not rejected by the test to form a confidence set. The asymptotic coverage of such a set will be at the declared level no matter the strength of identification.

\subsection{Robust AR}\label{sec:AR}

The AR test was developed  as an identification robust test in IV settings  with a small number of instruments. It uses the exogeneity assumption ($\E[e_i|Z_i]=0$), but not relevance ($\pi\neq 0$). The idea behind  the AR test, is that in order to test $H_0:\beta=\beta_0$, one tests whether the  implied errors $e(\beta_0)=Y-\beta_0X$, which coincide with the true structural errors for the correct value of $\beta_0$, are correlated with $Z$ in the sample.
The AR test statistic is  
$e(\beta_0)'Z\Sigma^{-1}Z'e(\beta_0)$, where
$\Sigma$ is the covariance matrix of $e'Z$ or a good estimate of it. Under the null, this test statistic has an asymptotic $ \chi^2_K$ distribution.  Under the further assumption of homoskedasticity of $e_i$ the statistic  reduces to  $\frac{1}{\hat\sigma^2}e(\beta_0)'Z(Z'Z)^{-1}Z'e(\beta_0)=\frac{1}{\hat\sigma^2}e(\beta_0)'P_Ze(\beta_0)$.

The AR test introduced in settings with a small number of instruments performs poorly in settings with a large number of instruments because the limit null distribution $\chi^2_K$ does not provide an accurate approximation. Notice that if a large number of instruments is modeled as $K\to\infty$, then the prescribed limit null distribution $\chi^2_K$ drifts to infinity. This issue is tightly related to the observation that under the null the test statistic  has a non-zero mean 
$$\E \left[e(\beta_0)'P_Ze(\beta_0)\right]=\sum_{i=1}^NP_{ii}
   \E e_i^2.$$
This coincides with the issue that the expected value of the  TSLS objective function is not minimized at the true parameter value $\beta_0$, contributing to its inconsistency when the number of instruments is large.  Therefore, the idea of jack-knifing or diagonal removing can be similarly applied to the original AR test statistic \citep{crudu_inference_2021,mikusheva_inference_2022}.  
 The infeasible JIV (or leave-one-out) AR statistic is defined as
\[
AR_{0}(\beta_{0})=\frac{1}{\sqrt{K\Phi_{0}}}\sum_{i\neq j}e_{i}(\beta_{0})P_{ij}e_{j}(\beta_{0}),
\]
where the normalizing factor $\Phi_{0}=\frac{2}{K}\sum_{i\neq j}P_{ij}^{2}\sigma_{i}^{2}\sigma_{j}^{2}$ is the variance of the quadratic form. Here and below we allow for a general form of heteroskedasticity where $\sigma_i^2=\E e_i^2$. Under minor assumptions like finite fourth moments of the errors and a well-balanced design  assumption,  the central limit theorem  for quadratic forms established in \cite{chao_asymptotic_2012} guarantees that 
 under $H_0:\beta=\beta_0$ we have $AR_0(\beta_0)\Rightarrow N(0,1)$.
It is worth pointing out that this theorem needs $K\to\infty$. The test rejects the null whenever we have a large positive value of the AR statistic.

\paragraph{Variance estimation.} In order to obtain a feasible test of an asymptotically correct size, one needs to estimate the variance $\Phi_{0}$. A good estimator should be consistent under the null and should allow for a general form of heteroskedasticity. Ideally, it should also ensure a good power under  alternatives. 

\cite{crudu_inference_2021}  proposed using the squared  implied errors  $\widehat{\sigma}_{i}^{2}=e_{i}^{2}(\beta_{0})$ as an unbiased proxy for $\sigma_i^2$ and the corresponding variance estimator is defined as
$$\widehat\Phi_{1}=\frac{2}{K}\sum_{i\neq j}P_{ij}^{2}\widehat\sigma_{i}^{2}\widehat\sigma_{j}^{2}.$$
It is very easy to show that under the null $\widehat\Phi_{1}$ is consistent for $\Phi_0$, and thus the test using $\widehat\Phi_{1}$  has an asymptotically correct size under a general form of heteroskedasticity. However, such a test may have low power against distant alternatives. Namely, if the true $\beta$ is very different from $\beta_0$, then $e(\beta_0)$ differs from structural errors $e$ by a potentially large predictable component. Squaring the implied errors would drastically overestimate the variances of error terms and may produce unnecessarily large values for $\widehat\Phi_{1}$. This may lead to a significant power loss especially at large deviations of the postulated $\beta_0$ from the true $\beta$.

Since the difference between the implied errors $e(\beta_0)$ and structural errors $e$ is predictable, one may residualize the
implied error before squaring.  Denote $M_Z=I-P_Z$ the projection matrix and let $M_i$ be its $i$th row. Even under the null, the squared residualized
error is biased:  $\E\left(M_{i}\mathbf{e}(\beta_{0})\right)^{2}\neq\sigma_{i}^{2}$. This is because the squared residual contains not only
the squared error $e_i$ but also the square of the regression estimation mistake. The latter can
be large when the number of regressors $K$ is large.

To construct an unbiased estimator for $\Phi_0$ under the null,  in \cite{mikusheva_inference_2022} we suggested the following estimator:
\[
\widehat{\Phi}_2=\frac{2}{K}\sum_{i\neq j}\frac{P_{ij}^{2}}{M_{ii}M_{jj}+M_{ij}^{2}} \left[e_{i}(\beta_{0})M_{i}e(\beta_{0})\right]\left[e_{j}(\beta_{0})M_{j}e(\beta_{0})\right].
\]
Our idea is based on the ``cross-fit'' variance proxies $
\widehat{\sigma}_i^2=\frac{1}{1-P_{ii}}e_i(\beta_0)M_ie(\beta_0),
$ proposed by 
 \cite{newey_cross-fitting_2018} and \cite{kline_leave-out_2020}. These proxies are  unbiased under the null: 
$
\E\widehat{\sigma}_i^2=\sigma_i^2.
$
However, since the normalizing factor  $\Phi_0$ is   quadratic  in $\sigma_i^2$ and the proxies for the variances of errors with different indexes (i.e. $\widehat\sigma_i^2$ and $\widehat\sigma_j^2$)   depend on the same sample, one can show that
$$\E\left[(e_iM_ie)(e_jM_je)\right]=(M_{ii}M_{jj}+M_{ij}^2)\sigma_i^2\sigma_j^2,
$$
and we can therefore obtain an unbiased estimator of $\Phi_0$ by a simple re-weighting of summands.

Another alternative estimator for $\Phi_0$ was proposed in
\cite{anatolyev_testing_2021}. Their idea is to create an unbiased proxy for $\sigma_i^2\sigma_j^2$ under the null by creating a product of four uncorrelated terms using a ``leave-three-out'' estimator. They suggested using $\widehat \Phi_3=\frac{2}{K}\sum_{i\neq j}P_{ij}^{2}\widehat{\sigma_{i}^{2}\sigma_{j}^{2}}$ with
$$\widehat{\sigma_i^2\sigma_j^2}=e_i(\beta_0)e_j(\beta_0)\sum_{k} \tilde{M}_{ik,-(ij)}e_k(\beta_0)[e_j(\beta_0)-Z_j'\widehat\delta_{-(ijk)}],
$$
where $\widehat\delta_{-(ijk)}$ are coefficient estimates from regressing $e(\beta_0)$ on $Z$ leaving three observations ($i,j,k$) out, while $\tilde{M}_{ik,-(ij)}$ is an element of the projection matrix leaving out $i$ and $j$.
There are explicit formulas for  leave-(one/two/three)-out projections available; however, the numerical complexity of implementing $\widehat \Phi_3$ is higher than the other two estimators. 

As for their theoretical properties, all three estimators of $\Phi_0$ are unbiased and consistent under the null. When the true value  $\beta=\beta_0+\Delta$ differs from the hypothesized value $\beta_0$, $\widehat \Phi_2$ and $\widehat \Phi_3$ are also consistent under local alternatives. In \cite{mikusheva_inference_2022} we derived the power function for the infeasible AR  uniformly over a set of local alternatives 
\[
AR_0(\beta_{0})\Rightarrow\Delta^{2}\frac{\mu^{2}}{\sqrt{K\Phi_{0}}}+\mathcal{N}(0,1).
\]
This is also the theoretical description of the power functions for AR using $\widehat\Phi_2$ or $\widehat\Phi_3$. One can show numerically that using $\widehat\Phi_1$ leads to a power loss.

\begin{figure}
\begin{centering}
\subfloat[sparse,$\frac{\mu^2}{\sqrt{K}}=2.5$]{\centering{}\includegraphics[scale=0.35] {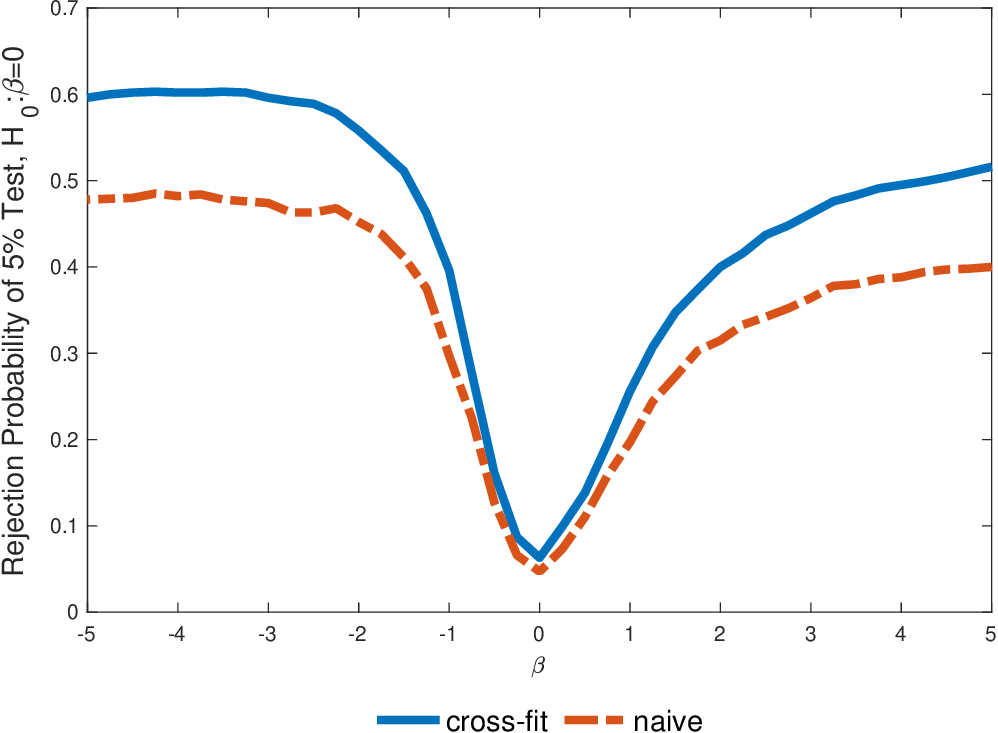}}\subfloat[dense,$\frac{\mu^2}{\sqrt{K}}=2.5$]{\centering{}\includegraphics[scale=0.35]{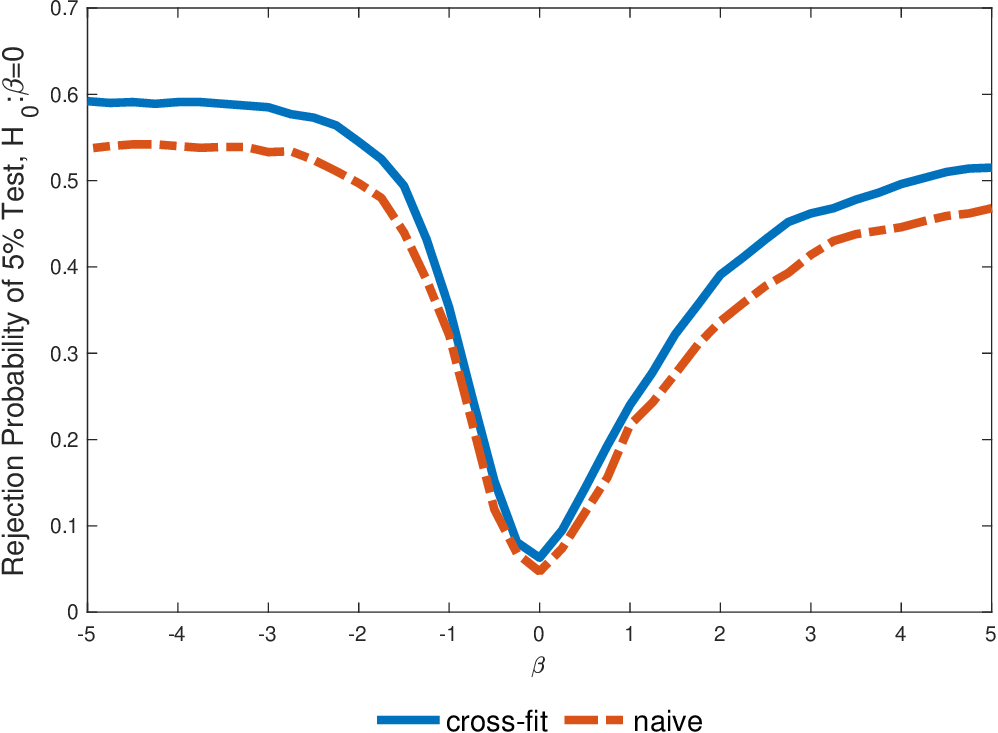}}
\par\end{centering}
\caption{Power curves for the leave-one-out AR tests with $\widehat\Phi_1$ (red dashed) and  $\widehat\Phi_2$ (blue solid)   variance estimators under sparse vs. dense first stage. Instruments are $K=40$ balanced group indicators, $N=200$, based on 1,000 simulations.}\label{fig- power of AR}
\end{figure}

\paragraph{Small-scale simulations.} We simulate the data according to a homoscedastic linear IV model (\ref{eq: iv model}) with a linear first stage $\Pi_i=\Pi'Z_i$ and true structural parameter $\beta=0$. The sample size is $N=200$. We divide the sample into $K=40$ equal groups, and define the instruments to be the group indicators. We simulate two designs with varying levels of sparsity. In the  sparse first stage we set one large coefficient $\pi_K=2$ and   $\pi_{k}=0.001$ for all $k<K$. This is the setting where one instrument contains almost all information. The dense first stage has homogeneous first stage coefficients  $\pi_{k}=0.316$ for all $k=1,\dots,K$. Identification strength is held the same at $\frac{\mu^2}{\sqrt{K}}=2.5$ for both designs.    The error terms $(e_{i},v_{i})$ are drawn i.i.d. from a Gaussian distribution with mean zero, unit variances, and correlation  $\rho=0.2$.  For each simulation draw we perform the leave-one-out AR tests using either $\widehat\Phi_1$ (red dashed line) or $\widehat\Phi_2$ (blue solid line). The resulting power curves are reported on Figure \ref{fig- power of AR}.

One can make two observations based on Figure \ref{fig- power of AR}. First, that the settings considered are  cases of weak identification, as the power curves stabilize on the level well below 1. Second, there is a significant power loss due to using a naive estimator for the scale estimator $\widehat\Phi_{1}$. The usage of estimator $\widehat\Phi_{2}$ is preferred from a power perspective.

\subsection{Robust LM}

In over-identified settings with a small number of instruments, the AR test is known to be asymptotically inefficient if identification is strong. The main reason is that under strong identification the data contains a lot of information about the optimal instrument, which the AR test completely ignores. The solution has been an identification robust modification of the Lagrange Multiplier (LM) test, known as the KLM test. 

An alternative modification to the LM test can make it robust to weak identification when there are many instruments.  The  LM test aims to construct  the most powerful combination of instruments and then conduct the AR test with the single instrument. The original LM  statistic is based on the linear combination $e'(\beta_{0})Z\widehat{\pi}=e'(\beta_{0})P_ZX$ because under homoskedasticity the optimal instrument is $Z\pi$. However, similar to the bias of the TSLS, in the setting of many instruments, the original LM statistic is poorly centered due to the correlation between the first stage $\widehat\pi$ and the structural error. As before, the problem can be solved by removing the diagonal.
The infeasible leave-one-out  LM statistic is
\[
LM^{1/2}(\beta_{0})=\frac{1}{\sqrt{K\Psi}}\sum_{i\neq j}e_{i}(\beta_{0})P_{ij}X_{j},
\]
where the normalization scalar is 
$$\Psi =\frac{1}{K}\sum_{i=1}^{N}(\sum_{j\neq i}P_{ij}X_{j})^{2}\sigma_{i}^{2}+\frac{1}{K}\sum_{i=1}^{N}\sum_{j\neq i}P_{ij}^{2}\gamma_{i}\gamma_{j},
$$
with
$\sigma_i^2=\E e_i^2, \gamma_i=\E[X_ie_i]$.
\cite{matsushita_otsu_2022}  proposed this test statistic and showed that under minor technical conditions, including the assumption that $K\to\infty$  as $N\to\infty$, under $H_0:\beta=\beta_0$ we have  $LM^{1/2}(\beta_0)\Rightarrow N(0,1)$. The LM test rejects 
when $\left|LM^{1/2}(\beta_{0})\right|$ is large (two-sided rejection).

As before, in order to implement this test one needs an estimator of $\Psi$.
Similar to \cite{crudu_inference_2021} in the case of the AR, \cite{matsushita_otsu_2022} suggested using the squared implied errors as proxies for variances under the null. Specifically, their proposed estimator is 
$$
\widehat{\Psi}_1  =\frac{1}{K}\sum_{i}\widehat{\sigma}_{i}^{2}(\sum_{j\neq i}P_{ij}X_{j})^{2}+\frac{1}{K}\sum_{i}\sum_{j\neq i}P_{ij}^{2}\widehat{\gamma}_{i}\widehat{\gamma}_{j},
$$
where $\widehat{\sigma}_{i}^{2}=e_{i}^{2}(\beta_{0})$ and $\widehat{\gamma}_{i}=X_{i}e_{i}(\beta_{0})$. \cite{matsushita_otsu_2022} showed that their estimator
$\widehat{\Psi}_1$ is consistent for $\Psi$ under the null ($H_0:\beta=\beta_0$), and thus, the feasible LM with $\widehat\Psi_1$ has the correct asymptotic size. However, for similar reasons as  in the case of the AR, the estimator $\widehat\Psi_1$ may lead to power losses under alternatives because $\widehat{\sigma}_{i}^{2}$ contains a large predictable part and overstates the variances $\sigma_i^2$.

In the current paper we propose a novel variance estimator $\widehat\Psi_2$ using ideas similar to \cite{mikusheva_inference_2022}:
$$
\widehat\Psi_2=\frac{1}{K}\sum_{i}\frac{e_{i}M_{i}e}{M_{ii}}(\sum_{j\neq i}P_{ij}X_{j})^{2}+\frac{1}{K}\sum_{i}\sum_{j\neq i}\widetilde{P}_{ij}^{2}X_{i}M_{i}eX_{j}M_{j}e.
$$
 Here we use 
$\widehat\sigma_i^2=\frac{e_{i}M_{i}e}{M_{ii}}$ , an unbiased proxy for $\sigma_i^2$,
 $\widehat\gamma_i=X_iM_ie$ , a  proxy for $\gamma_i$, and
 re-weighting $\widetilde P_{ij}^2=\frac{P_{ij}^2}{M_{ii}M_{jj}+M_{ij}^2}$ to correct for correlation in proxies.

We also establish a new theoretical result showing that under both  the null and   local alternatives this estimator $\widehat\Psi_2$ is consistent under a general form of heteroskedasticity. For that we need the following assumptions.

\begin{assumption}\label{assumption- few controls}
{\ }
\begin{itemize}
\item[(i)] $P_Z$ is an $N\times N$ projection matrix of rank $K$, $K\to\infty $ as $N\to\infty$ and there exists a constant $\delta$ such that $\max_iP_{ii}\leq \delta<1$;
\item[(ii)] Errors $\varepsilon_i=(e_i,v_i)', i=1,...,N$ are independent with $\E\varepsilon_i=0,\max_i\E\|\varepsilon_i\|^6<\infty$, and for some positive constants $c^*$ and $C^*$ that do not depend on $N$
$$
c^*\leq\min_i\min_x\frac{x'Var(\varepsilon_i)x}{x'x}\leq\max_i\max_x\frac{x'Var(\varepsilon_i)x}{x'x}\leq C^*.
$$
\end{itemize}
\end{assumption}

Assumption \ref{assumption- few controls} is a very straightforward generalization of assumptions needed for validity of robust AR test under many instruments. Specifically we only added the moment conditions on the first stage error $v_i$ and positive-definiteness of the error covariance matrix.

\begin{theorem}\label{thm- consistency of LM}
Let Assumption 1 hold, and  $\frac{\pi^{\prime}Z'Z\pi}{K^{2/3}}\to0$ as $N\to\infty$,
then
\begin{itemize}
    \item[(1)]  if $\beta=\beta_{0}$,
we have $\frac{\widehat\Psi_2}{\Psi}\to^{p}1$ as $N\to\infty;$
\item[(2)] if $\beta=\beta_{0}+\Delta$ such that $\Delta\cdot\frac{\pi'Z'Z\pi}{K}\to 0$,
we have $\frac{\widehat\Psi_2}{\Psi}\to^{p}1$ as $N\to\infty.$
\end{itemize}
\end{theorem}

Part (1) of Theorem \ref{thm- consistency of LM} gives consistency under the null hypothesis, and thus implies that the leave-one-out LM test using $\widehat\Psi_2$ has an asymptotically correct size. Part (2) addresses the consistency under local alternatives and guarantees that the power curves of the LM test using our proposed estimator $\widehat\Psi_2$ are the same as those of the infeasible LM test. Specifically, the power function of the infeasible leave-one-out LM test is described by the following convergence 
uniformly over the alternatives $\beta=\beta_0+\Delta$: 
\[
LM^{1/2}\Rightarrow\Delta\frac{\mu^{2}}{\sqrt{K\Psi}}+\mathcal{N}(0,1).
\]
Notice that $\Delta$ can be both positive or negative, thus, if one uses $LM^{1/2}(\beta_0)$ statistics then they can employ the standard Gaussian critical values with  two-sided rejection. Alternatively, one may use the squared statistic and the 95-percentile of $\chi^2_1$ distribution. Another observation is that the proposed LM test is consistent for fixed alternatives
as soon as $\mu^{2}/\sqrt{K}\rightarrow\infty$.

\begin{figure}
\begin{centering}
\subfloat[sparse,$\frac{\mu^2}{\sqrt{K}}=2.5$]{\centering{}\includegraphics[scale=0.35] {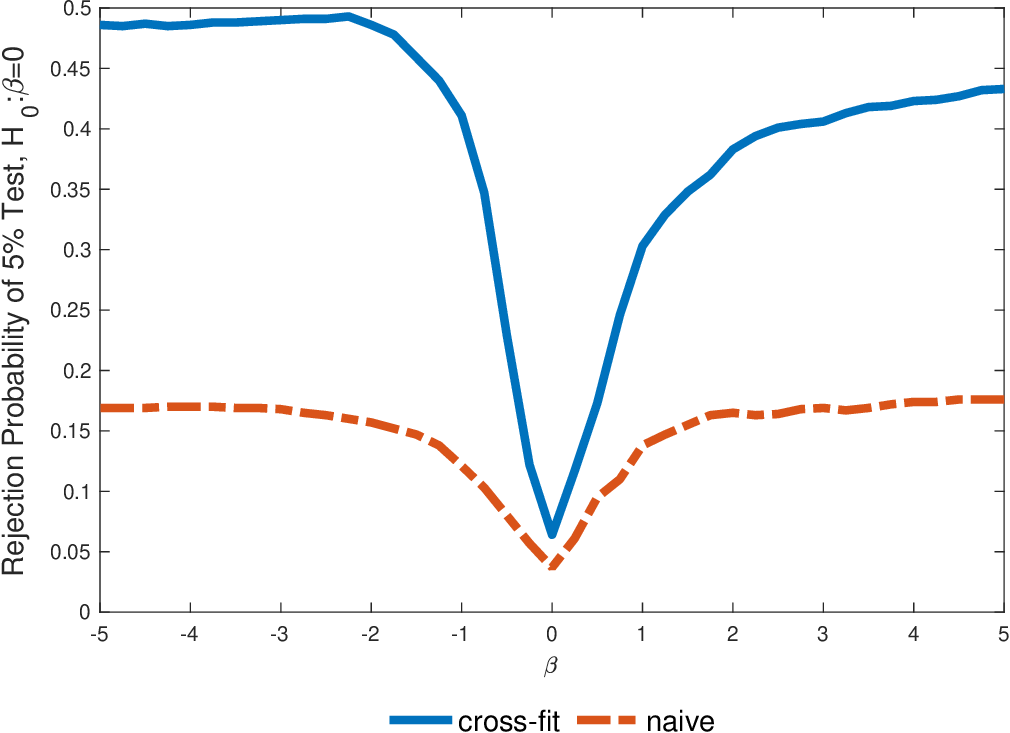}}\subfloat[dense,$\frac{\mu^2}{\sqrt{K}}=2.5$]{\centering{}\includegraphics[scale=0.35]{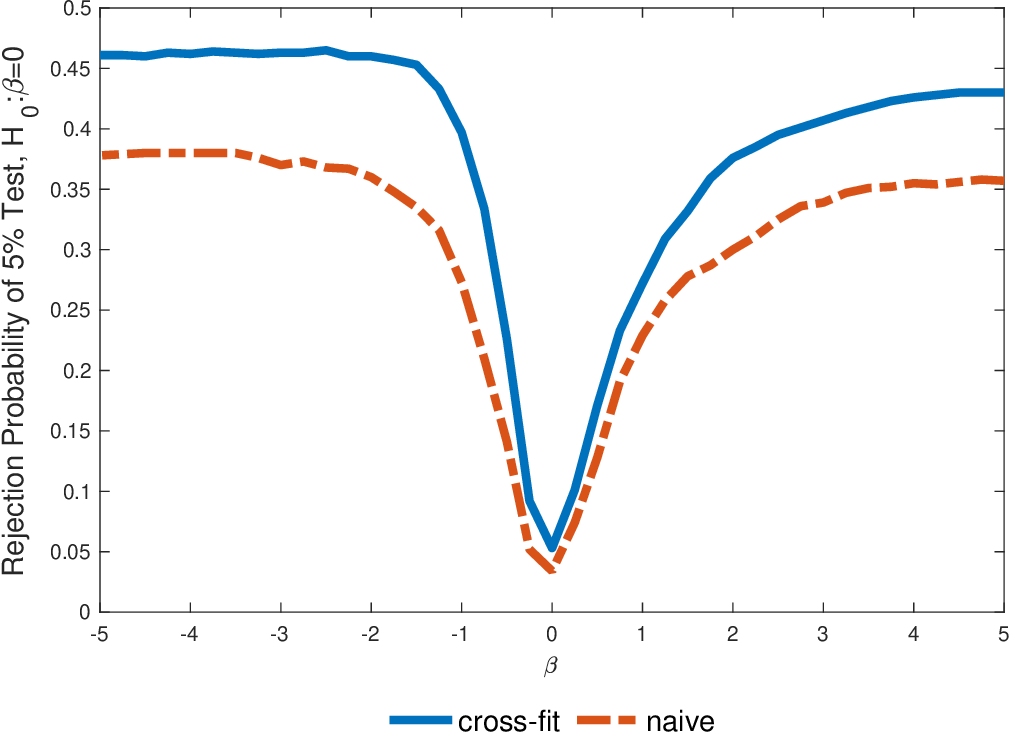}}
\par\end{centering}
\caption{Power curves for the leave-one-out LM  with $\widehat\Psi_1$ (red dashed) and  $\widehat\Psi_2$ (blue solid)   variance estimators under sparse vs. dense first stage. Instruments are $K=40$ balanced group indicators, $N=200$.}\label{fig- LM power}
\end{figure}

\paragraph{Small-scale simulation (continued).}
We repeat the same simulation design as in Section~\ref{sec:AR}, where Assumption \ref{assumption- few controls} is trivially satisfied.  We calculate the power curves for the leave-one-out LM test using $\widehat\Psi_1$ (red dashed line) and $\widehat\Psi_2$ (blue solid line) estimates of the scale. They are reported in Figure \ref{fig- LM power}. Here the power loss due to usage of the naive estimate of the scale $\widehat\Psi_1$ is extremely pronounced, especially in the sparse design. Another observation is that in both designs, contrary to the conjecture that LM is more efficient than AR, the leave-one-out AR test has higher power than the leave-one-out LM.  This is not a universal observation and we discuss their trade-offs in the next sub-section.

\paragraph{Empirical example (Angrist and Krueger, 1991).}
\begin{table}
\begin{centering}
\small{
\begin{tabular}{cccccc}
\hline
 &FF &$\widetilde{F}$ & JIVE-t-test & Robust AR & Robust LM\tabularnewline
\hline
\hline
180 instruments & 2.4 &13.4 & {[}0.066,0.132{]} & {[}0.008,0.201{]} & {[}0.067,0.135{]}\tabularnewline
\hline
1530 instruments  & 1.3 & 6.2 & {[}0.024,0.121{]} & {[}-0.047, 0.202{]} & {[}0.022,0.127{]}\tabularnewline
\hline
\end{tabular}
}
\par\end{centering}

\caption{\label{tab:AK91}Robust and non-robust confidence sets in \cite{AngristKrueger1991} } 

{\footnotesize{\emph{Notes: }Results on pre-tests for weak identification and the confidence sets based on the JIVE t-test, the leave-one-out AR and the leave-one-out LM for IV  specification underlying Table VII
Column (6) of \cite{AngristKrueger1991}.  The confidence sets are constructed via analytical test inversion. }} 

\end{table}
We return to the empirical example of \cite{AngristKrueger1991} and report the robust confidence sets obtained by inverting the leave-one-out  AR and LM  tests robust to many weak instruments in Table \ref{tab:AK91}. We note that all confidence sets are finite intervals and are somewhat informative, though the leave-one-out AR confidence is significantly wider than the other two. This is mostly due to the fact we established before that the identification seems to be strong in \cite{AngristKrueger1991}. We may also notice that the leave-one-out LM confidence set is nearly identical to the JIVE-Wald confidence set. This is a reflection of the fact that the leave-one-out LM test is asymptotically equivalent to the JIVE $t$-test under strong identification. Unlike   the JIVE $t$-test, which fails to controls size under weak identification, the leave-one-out LM test is fully robust to weak identification. Thus, we  recommend using the leave-one-out LM test as a default option  without a pre-test for weak identification. Another observation is that the case with 1,530 instruments is less informative and produces wider confidence sets for all three test statistics  than when only 180 instruments are used. 

\subsection{Combination tests}

In the empirical example of \cite{AngristKrueger1991} we have seen that the leave-one-out LM test produces much shorter confidence sets than the robust leave-one-out AR test. At the same time our simulation exercise showed the opposite ordering of power.  This raises the question of power comparison between these two  tests. The answer  is that neither of the two tests dominates the other. As  mentioned before, the power curves for the infeasible tests (or for  the feasible tests with our proposed estimators  $\widehat\Phi_{2}$ or $\widehat\Psi_{2}$ for the normalizing factors) 
under the alternative $\beta=\beta_0+\Delta$  and when $\frac{\mu^2}{K}\to 0$ can be characterized by:
\begin{align*}
LM^{1/2}\Rightarrow\Delta\frac{\mu^{2}}{\sqrt{K\Psi}}+\mathcal{N}(0,1),
\\
AR \Rightarrow\Delta^2\frac{\mu^{2}}{\sqrt{K\Phi}}+\mathcal{N}(0,1).
\end{align*}

These power curves imply that when a setting is strongly identified in the sense of \cite{mikusheva_inference_2022}, that is, when $\frac{\mu^{2}}{\sqrt{K}}\to\infty$, then both the leave-one-out AR and the leave-one-out LM are asymptotically consistent for fixed alternatives $\beta$.
Under strong identification the two tests have different sets of
local alternatives, namely, alternatives with asymptotically non-trivial probability of detection. Specifically,
for  the leave-one-out AR test the set of local alternatives is $\{\Delta:\frac{\Delta{}^{2}\mu^{2}}{\sqrt{K}}= C\}$ i.e.,
$|\Delta|\propto\sqrt{\frac{\sqrt{K}}{\mu^{2}}}$, while for the leave-one-out LM  test it is
 $\{\Delta:\frac{|\Delta|\mu^{2}}{\sqrt{K}}= C\}$ i.e., $|\Delta|\propto\frac{\sqrt{K}}{\mu^{2}}$. So, we observe that under strong identification the leave-one-out 
 AR test has a slower speed of detection than the leave-one-out LM. This suggests that when identification is strong the leave-one-out LM will tend to produce shorter confidence sets.

However, when identification is weak  ($\frac{\mu^{2}}{\sqrt{K}}$ is bounded), there is no consistency for fixed alternatives, and the confidence sets tend to be non-shrinking. In such settings, the leave-one-out AR will tend to have higher power for distant alternatives and would be preferable to use. The absence of ordering between AR and LM tests in terms of power under weak identification has been known in a case with fixed number of instruments, our findings extend this to many instruments case.

These considerations suggest that one  may want to combine the robust test statistics in some optimal way based on  the strength of identification. For example, one may do a switch or ``soft switch'' between the leave-one-out AR and the leave-one-out LM statistics depending on the size of $\widetilde F$. In cases with small number of instruments, \cite{Andrews2016} found the conditional likelihood ratio (CLR) test of \cite{Moreira2003} has very good power properties because it can be considered as a robust test that implements a soft switch. A recent paper by \cite{ayyar_conditional_2022} suggests how to construct a weak instrument robust analog of the CLR test when the number of instruments is large. A recent paper by \cite{lim_conditional_2022} searches for an optimal linear combination test that optimizes a minimax criterion that was first proposed in \cite{Andrews2016}. The resulting test statistic is a weighted average of the leave-one-out AR and the leave-one-out LM.

\section{Allowing for many exogenous regressors}\label{section- covariates}
To illustrate the issue arising from many instruments, previously we simplified the exposition by assuming no exogenous regressors (controls) in the structural equation. However, this assumption is unrealistic in practice. In models with many instruments, it is typical to also have many exogenous regressors. While the specification underlying Table \ref{tab:AK91} contains a relatively small number of covariates (71), a more modern approach would increase the number of covariates dramatically. For example, in settings like \cite{AngristKrueger1991} once we use interactions of baseline instruments with covariates as instruments, it is common to include those covariates as exogenous regressors as well. More broadly, in practice many instrumental variables are only valid after conditioning on additional covariates.  ``Saturated'' specifications that control for covariates nonparametrically can ensure proper interpretation of the IV estimate under the LATE framework of \citet{ImbensAngrist1994}. A ``saturated'' specification includes rich interactions between the instrument and the covariates, giving rise to many instruments and many exogenous regressors, as recently discussed in  \cite{sloczynski_when_2020} and \cite{Blandhol2022}.

To analyze the impact of many exogenous regressors, we consider the following model:
\begin{equation}\label{eq: iv model}
\left\{ \begin{array}{c}
          Y_i=\beta X_i + \gamma' W_i + e_i ,\\
          X_i= \pi' Z_i+ \delta' W_i + v_i,
        \end{array}
\right.
\end{equation}
for $ i=1,..., N.$   This is a linear IV regression with a scalar outcome $Y_i$, an endogenous scalar regressor $X_i$, a $K_Z\times 1$ vector of instrumental variables  $Z_i$ and a  $K_W\times 1$ vector of exogenous regressors $W_i$. The main assumption is $\E[e_i|Z_i, W_i]=\E[v_i|Z_i,W_i]=0$. We  are interested in estimation of and statistical inference about $\beta$, and treat parameters $\pi, \gamma,\delta$  as nuisance parameters.

\paragraph{New challenges of estimation with many instruments and many exogenous regressors.}
The TSLS estimator of $\beta$ in this case is equivalent to the TSLS in a model that first partials out $W$. Let us introduce a projection matrix $M_W=I-W(W'W)^{-1}W'$. Denote $Y^{\perp}=M_WY$, $X^{\perp}=M_WX$ and $Z^{\perp}=M_WZ $ to be the outcome variable, endogenous regressor and instruments with the exogenous regressors partialled out.  Finally denote $P^{\perp}=P_{Z^{\perp}}=Z^{\perp}\left((Z^{\perp})'Z^{\perp}\right)^{-1}(Z^{\perp})'$ to be the projection matrix based on the residualized instruments. 
Then the TSLS estimator of $\beta$ is
$$
\widehat\beta_{TSLS}=\frac{(X^\perp)'P^{\perp}Y^\perp}{(X^\perp)'P^{\perp}X^\perp}.
$$
As we have seen in the case without exogenous regressors, the TSLS estimator is very biased when $K_Z$ is large. We should expect a similar issue in the case with many instruments and many exogenous regressors. One solution in the case without exogenous regressors is to remove a diagonal from the projection matrix, so, a natural though  naive approach is the following estimator 
$$
\widehat\beta_1=\frac{\sum_{i\neq j}X^\perp_iP^{\perp}_{ij}Y^\perp_j}{\sum_{i\neq j}X^\perp_iP^{\perp}_{ij}X^\perp_j}.
$$
This estimator was proposed in \cite{ackerberg2009improved} under the name improved JIVE (IJIVE).
Here we first partial out exogenous regressors, and then use a JIVE- type estimator without exogenous regressors that removes a diagonal from the projection on the instruments. \cite{evdokimov2018inference} showed that this approach does not quite work and $\widehat\beta_1$ tends to have large biases.  Indeed,
$$
\widehat\beta_1-\beta=\frac{\sum_{i\neq j}X^\perp_iP^{\perp}_{ij}e^\perp_j}{\sum_{i\neq j}X^\perp_iP^{\perp}_{ij}X^\perp_j}, \mbox{  where  } e^\perp=M_We.
$$
We can show that the numerator has a non-trivial mean. Let us denote $M_{W,ij}$ as elements of $M_W$ and use an observation that $M_WP^\perp=P^\perp M_W=P^\perp$:
\begin{align*}
\E\sum_{i\neq j}X^\perp_iP^{\perp}_{ij}e^\perp_j=\sum_k\sum_{i\neq j}
\E[v_ke_k]M_{W,ki}P^\perp_{ij}M_{W,kj}= \sum_{k}
\E[v_ke_k] P^\perp_{kk}(1-M_{W,kk}).
\end{align*}
The last expression is non-trivial, since on average $1-M_{W,kk}$ is $\frac{K_W}{N}$, while the trace of $P^{\perp}$ is $K_Z$. Thus,  under homoskedasticity, if the values of $P^\perp_{kk}$ are uncorrelated with $M_{W,kk}$, we should expect the last expression to be $\sigma^2\frac{K_WK_Z}{N}$. Removing the diagonal from $P^\perp$ has not solved the many instruments issue here, because the procedure of partialling out introduces the dependence across residuals so that  $e^\perp_i$ are not independent across $i$. 

Another suggestion is to write the numerator of the TSLS as $(X^\perp)'P^{\perp}Y^\perp=X'P^\perp Y$, which is due to $M_WP^\perp M_W=P^\perp$, and to remove the diagonal from $P^\perp$. That is, one may propose the following estimator:
$$
\widehat\beta_2=\frac{\sum_{i\neq j}X_iP^{\perp}_{ij}Y_j}{\sum_{i\neq j}X_iP^{\perp}_{ij}X_j}.
$$
This suggestion does not work either but for a different reason. The operator $P^\perp$ has a property that it projects out the exogenous regressors, namely $P^\perp W=0$, but the same operator without the diagonal does not have this property: $\sum_{j\neq i}P^\perp_{ij}W_j'\neq 0$. Thus
$$
\widehat\beta_2-\beta=\frac{\sum_{i\neq j}X_iP^{\perp}_{ij}W_j'\gamma+\sum_{i\neq j}X_iP^{\perp}_{ij}e_j}{\sum_{i\neq j}X_iP^{\perp}_{ij}X_j}.
$$
The term in the numerator $\sum_{i\neq j}X_iP^{\perp}_{ij}W_j'\gamma=-\sum_{i}X_iP^{\perp}_{ii}W_i'\gamma$ corresponds to a bias. Since the average value of $P_{ii}^\perp$ is $\frac{K_Z}{N}$, this term is (approximately) the fraction of the bias that would arise if one does not include exogenous regressors in the regression (omitted variable bias).

To summarize, the challenges from both many instruments and many exogenous regressors are a counterplay of two needs: the need to partial out exogenous regressors and the need to remove the diagonal. An ideal estimator takes the  form of
$$
\widehat\beta_3=\frac{X'A Y}{X'AX},
$$
where a $N\times N$ matrix $A$ has the following properties: (i) $AW=0$ (partialing out property) and (ii) $A_{ii}=0$ for all $i$ (zero diagonal property). Matrix $A$ can be constructed using $Z$ and $W$, with some preferences for a matrix close to $P^\perp$.

A recent paper by \cite{chao_jackknife_2022} suggests a matrix $A$ of the following form $A=M_W(P^\bot -D_\theta)M_W $  where $D_\theta$ is a diagonal matrix with diagonal elements  $\theta_1,...,\theta_N$  selected in such a way that $A$ has zero diagonal. 
 \cite{chao_jackknife_2022}  showed that such $\theta_i$'s can be found in a well-balanced design (when $\min_i M_{W,ii}>1/2$) and provided a proof of consistency and asymptotic Gaussianity  of estimator $\widehat\beta_3$ under some assumptions. \cite{kolesa2013estimation} suggested an alternative matrix $A$ for bias correction, yielding an estimator with a similar form to $\widehat\beta_3$. For this paper, we focus on the matrix $A$ proposed by \cite{chao_jackknife_2022} to construct tests robust to many instruments/exogenous regressors.

\paragraph{Robust AR.} Following the ideas stated in the previous sections, we create a test for $H_0:\beta=\beta_0$  robust to many instruments/exogenous regressors using an AR-type statistic
$$
AR_W(\beta_0)=\frac{1}{\sqrt{K_Z\widehat\Phi_W}}(Y-\beta_0X)'A (Y-\beta_0X),
$$
where we propose a novel estimator for the normalization factor
\begin{align}\label{eq: def PhiW}
    \widehat\Phi_W=\frac{2}{K}\sum_{i,j}\frac{A_{ij}^2}{M_{ZW,ii}M_{ZW,jj}+M_{ZW,ij}^2}\widehat\sigma_i^2\widehat\sigma_j^2.
\end{align}
Here
$\widehat\sigma_i^2=\sum_{k}M_{ZW,ik}(Y_i-\beta_0X_i)(Y_k-\beta_0X_k)$, where $M_{ZW,ij}$ are elements of $M_{ZW}$, a projection matrix orthogonal to both $Z$ and $W$. This proposal is similar to the $\widehat\Phi_2$ estimator in the case with no exogenous regressors. There is no consistent ``naive'' variance estimator (like $\widehat\Phi_1$) in the presence of exogenous regressors since under the null $Y_i-\beta_0X_i$ is not equal to the error but rather contains the predictable non-zero part $(\gamma-\beta_0\delta)'W_i$. This term squared  overstates the true variance $\sigma_i^2$ drastically. The ideas of cross-fit variance estimation are very useful here. We propose a test that rejects the null when $AR_W(\beta_0)$ exceeds the right $\alpha$- quantile of the standard normal distribution. We show this test has the correct size.

\begin{assumption}\label{assumption- many controls}
{\ } 
\begin{itemize}
\item[(i)] Projection matrices $M_W$ and $P^\perp$ are such that $\min_i M_{W,ii}>1/2$, all components of vector $\theta=(M_W\circ M_W)^{-1}diag(P^\perp)$ are non-negative; there exists a constant $\delta$ such that $\frac{P^\perp_{ii}}{M_{W,ii}^2}\leq \delta<1$, and the rank $K_Z$ of projection matrix $P^\perp$ grows to infinity when $N\to\infty$ ;
\item[(ii)] Errors $e_i, i=1,...,N$ are independent with $\E e_i=0,\max_i\E\|e_i\|^6<\infty$, and for some positive constants $c^*$ and $C^*$ that do not depend on $N$
$$
c^*\leq\min_i\E\|e_i\|^6\leq\max_i\E\|e_i\|^6\leq C^*.
$$
\end{itemize}
\end{assumption}
Assumption~\ref{assumption- many controls} (i) can be characterized as an assumption about a balanced design. In a case with no exogenous regressors ($M_W=I$) this assumption is equivalent to Assumption~\ref{assumption- few controls}~(i). Since inferences are done conditionally on $W$ and $Z$, Assumption~\ref{assumption- many controls}~(i) can be directly assessed in a specific dataset for each application.

\begin{theorem}\label{thm- AR many controls}
Let Assumption 2 hold in model (\ref{eq: iv model}), 
then under the true null hypothesis $H_0:\beta=\beta_0$ as $N\to\infty$ we have
$$AR_W(\beta_0)\Rightarrow N(0,1).$$
\end{theorem}

\paragraph{Remark.} We conjecture that an alternative AR type statistic can be constructed using estimator proposed in \cite{kolesa2013estimation} in the same way we constructed statistic $AR_W$ based on estimator of \cite{chao_jackknife_2022}. Additionally, constructing an LM statistic with asymptotically correct size under both many instruments and many controls should be straightforward  by combining the ideas stated in this section and Theorem \ref{thm- consistency of LM}.

\paragraph{Simulation study.} In order to assess the size property of the newly proposed robust test and to compare it with naive approaches to deal with both many instruments and many exogenous regressors in a realistic setting we calibrate the simulation to \cite{gilchrist_something_2016}.

\cite{gilchrist_something_2016} are interested  in estimating social spillovers from movie viewership, namely,  the effect of viewership from a movie's opening weekend on subsequent viewership. To identify the causal effect they use weather during the opening weekend as  set of exogenous instruments. The setting contains both a large number of instruments and a large number of exogenous regressors. The set of instruments includes 52 different measures of weather conditions around a movie theater, including temperature, indicators for snow/rain, precipitation, etc. The set of exogenous regressors includes indicators for calendar year, day of the week, week of the year, holidays, as well as weather conditions in periods for which subsequent viewership is measured. The number of exogenous regressors is relatively high in comparison to the dimension of the instruments, which provides an empirically relevant setting for showing that the original leave-one-out AR test might not be robust to many exogenous regressors, and the adjustment proposed in this paper is able to restore the correct size.  

In order to calibrate the simulation to \cite{gilchrist_something_2016}, we follow the simulation design proposed in \cite{angrist_machine_2022}. Specifically, we take the LIML estimator of the model from the original data as the ground truth for $\beta$. Let $\widehat{y}(W_{i})$ be the linear function in $W_i$ equal to the dependent variable's fitted value, after subtracting $\widehat{\beta}_{LIML}X_{i}$.  We set $\pi$ to be the first stage coefficients. 
We simulate the data from the model:
\begin{align*}
    \widetilde{Y}_{i}=\widehat{y}(W_{i})+\beta\widetilde{X}_{i}+\omega_{i}(\epsilon_{i}-1.5v_{i}), \\
    \widetilde{X}_{i}=\pi'Z_{i}^{\perp}+v_{i}.
\end{align*}  
Here  $\beta=0.6$, weights $\omega_{i}$ are the absolute values of the LIML residuals to mimic the heteroskedasticity of the data.  We perform 1,000 simulations, drawing $(v_{i},\epsilon_{i})$ independently from the standard normal distribution. After removing multi-collinearities, we have $K_{Z}=48$ and $K_{W}=119$
for a sample size $N=1,669$. We check that Assumption 2(i) holds in this setting.

We calculate the simulated size for our proposed AR test robust to many instruments and many exogenous regressors by comparing statistics $AR_W(\beta_0)$ with the upper 95\% quantile of the standard normal distribution. We also check two naive approaches to testing using the AR test by calculating statistics
\begin{align*}
AR_1(\beta_0)&=\frac{1}{\sqrt{K_Z\widehat\Phi_1}}\sum_{i\neq j}P^{\perp}_{ij}(Y_i^\perp-\beta_0X_i^\perp)'(Y^\perp_j-\beta_0X^\perp_j),
\\
AR_2(\beta_0)&=\frac{1}{\sqrt{K_Z\widehat\Phi_2}}\sum_{i\neq j}P^{\perp}_{ij}(Y_i-\beta_0X_i)'(Y_j-\beta_0X_j),
\end{align*}
where $\widehat\Phi_i$ are properly constructed estimators of the normalization factor using the cross-fit ideas stated in this paper. The results are reported in Table \ref{tab:simulation Wald size}.

\bigskip
\begin{table}[h]
\begin{centering}
\begin{tabular}{cccccc}
\hline
$N$ & $K_{Z}$ & $K_{W}$ &  size of & size of &  size of
\tabularnewline
 &  &    & $AR_1$  & $AR_2$ & $AR_W$\tabularnewline
\hline\hline 
1,669 & 48 & 119 &  11\% & 1.3\% & 5\%\tabularnewline
\hline
\end{tabular}
\par\end{centering}
\caption{\label{tab:simulation Wald size}   Simulation
results for size of different modifications of the AR tests with many instruments and many exogenous regressors. Simulation design mimics data from Gilchrist and Sands (2016). }
\end{table}
The results show that if the test statistics fail at any of the two tasks, either at removing the diagonal (as $AR_1$ does) or at partialling out the exogenous regressors (as $AR_2$) then the size may differ from the declared level. At the same time a properly constructed statistic that by construction performs both tasks, paired with the proper estimator of the normalization factor, controls size effectively even with a large number of both instruments and exogenous regressors. Similarly, the estimators $\hat{\beta}_1$ and $\hat{\beta}_2$ are badly biased whereas $\hat{\beta}_3$ is roughly centered around the truth $\beta=0.6$.
\bigskip
\begin{table}[h]
\begin{centering}
\begin{tabular}{cccccc}
\hline
$N$ & $K_{Z}$ & $K_{W}$ &  bias of & bias of &  bias of
\tabularnewline
 &  &    & $\hat{\beta}_1$  & $\hat{\beta}_2$ & $\hat{\beta}_3$\tabularnewline
\hline\hline 
1,669 & 48 & 119 &  74\% & 67\% & -0.2\%\tabularnewline
\hline
\end{tabular}
\par\end{centering}
\caption{\label{tab:simulation bias}   Average of $(\hat{\beta}-\beta)/\beta$ for different modifications of JIV estimators $\hat{\beta}$ under many exogenous regressors. Simulation design mimics data from Gilchrist and Sands (2016). }
\end{table}
\section{Conclusion and open questions}

The goal of this paper is to provide an overview for the statistical challenges surrounding estimation and inferences in a linear IV model with many instruments. We show that aside from the obvious benefits of bringing additional identifying information, many instruments come at a cost as one typically needs to estimate the optimal way to combine many instruments. If one has many not very informative instruments, then the uncertainty surrounding  the first stage estimation may produce significant biases of the TSLS, and even lead to an inconsistency.
 
We showcased one set of methods and ideas that allowed reliable estimation and inferences. One of the central ideas, jack-knifing or deleting a diagonal,  produces both new estimators with superior convergence properties  and new identification robust tests. 

We presented results established in the econometric literature that inform a coherent empirical strategy. Specifically, one may use  a pre-test for weak identification robust to heteroskedasticity presented in Section \ref{section: pre-test}, and depending on its results either use the JIV estimator based on the idea of removing the diagonal paired with its standard errors, or use any of the identification robust tests presented in Section \ref{section- robust tests}. This paper also establishes some new results including a version of the  LM test  robust to many weak instruments with a new variance estimator and a modification of the AR test robust to both many instruments and many exogenous regressors.
 
As a final word we wish to mention several open questions in this research area that  have a chance to be solved within the next few years and we hope this encourages some researchers looking for  next  project to tackle them. 

The first open question is related to the observation that a pre-test for weak identification is tightly related to an estimator one hopes to use and is formulated as whether one can trust a specific estimator with a  confidence set or a test based on it.  The current test for weak identification is created for the JIVE. However, there are results pointing out that  other estimators like the JIVE-LIML (or its heteroskedasticity-robust version) are more efficient than the JIVE under strong identification (\cite{hausman_instrumental_2012}). It is still an open question to establish a pre-test for reliability of the heteroskedasticity-robust JIVE-LIML. Along the same line of thoughts, currently there is no pre-test for any estimator that accommodates not just many instruments but also many exogenous regressors, though empirically there is an important need for such a pre-test.

This paper discussed in detail  one approach based on jack-knifing or diagonal removal. There are other influential ideas mentioned in Section \ref{section - bias of TSLS} related to instrument selection or construction of the optimal instrument using ML approaches. Those ideas seem very powerful and produce quite efficient estimators if the first stage can be well described by a model in which the selected ML technique produces a consistent estimator of the optimal instrument.  Unfortunately, the performance of an ML first stage is generally unknown if the first stage does not satisfy the assumptions of a model needed for  ML consistency. Based on simulation studies from \cite{angrist_machine_2022} we are pessimistic that the aforementioned techniques work well under many weak instruments. As has been shown in \cite{mikusheva_many_weak_2022} the conditions needed for consistency of an IV estimator using LASSO selection on the first stage depends in a significant way on the true sparsity of the true first stage model. It has also been suggested in  \cite{mikusheva_many_weak_2022} that using sample-split and cross-fit may be a powerful idea for breaking the dependence between two stages when ML techniques are used on the first stage. Unfortunately, currently there is no good technique to assess whether an IV estimator with some ML algorithm used in the first stage is reliable in any given data set.  One technical challenge for developing such methods is  understanding the asymptotic behavior of ML estimators in  settings where modeling assumptions needed for consistency of an ML algorithm (like sparsity) do not hold.

It is worth pointing out that this paper as well as a vast majority of research papers devoted to many and/or weak instruments are written for cross-sectional settings, while there is a large number of empirical settings using macroeconomic or financial data that can be labeled as many weak instruments.  
\cite{mikusheva_many_weak_2022} showcased a very clear need to develop inferences robust to many weak instruments in time series settings as well as the associated challenges.

 \bibliography{Weak_IV} 
\appendix

\section{Appendix with Proofs}
Let $C$  be a universal constant (that may be different in different lines but does not depend on $N$ or $K$).

\paragraph{Proof of Theorem  \ref{thm- consistency of LM}.} 
The assumptions laid out in this theorem are exactly the ones stated in Theorem 5 of \cite{mikusheva_inference_2022}.  The only difference is that since we focus on linear first stage that the residualization is complete and satisfies $\Pi'M\Pi=0$.
Part (1) of Theorem \ref{thm- consistency of LM} follows from Lemma S3.1 (a) from the Supplementary
Appendix to \cite{mikusheva_inference_2022}, which 
establishes the consistency of the scale parameter  $\widehat\Psi_2$
under the null hypothesis. 

Part (2) addresses the consistency of the scale parameter for local alternatives
$$ e(\beta_0)=e+\Delta v+\Delta \cdot Z\pi=\eta+\Delta \cdot Z\pi.
$$
Notice that $Me(\beta_0)=M\eta$ as the predicted part partials out completely. Define
\[\tilde\Psi_2=\frac{1}{K}\sum_{i}\frac{\eta_{i}M_{i}\eta}{M_{ii}}(\sum_{j\neq i}P_{ij}X_{j})^{2}+\frac{1}{K}\sum_{i}\sum_{j\neq i}\widetilde{P}_{ij}^{2}X_{i}M_{i}\eta X_{j}M_{j}\eta.\]
Part(1) of Theorem  \ref{thm- consistency of LM} shows that $\tilde\Psi_2/\Psi_2\to^p 1$
as long as $\Delta\to 0$.  Given the partialling out property we have:
\[
\widehat\Psi_2-\tilde \Psi_2=\frac{1}{K}\sum_{i}\frac{\Delta Z_{i}\pi M_{i}\eta}{M_{ii}}(\sum_{j\neq i}P_{ij}X_{j})^{2}.
\]
We now apply part (d) of Lemma S3.2 from the Supplementary
Appendix to \cite{mikusheva_inference_2022} noticing that $w_i=M_{ii}Z_i\pi$ in notations of that lemma:
\[
\frac{1}{K}\sum_{i}\frac{\Delta Z_{i}\pi M_{i}\eta}{M_{ii}}(\sum_{j\neq i}P_{ij}X_{j})^{2}- \frac{2}{K}\sum_{i}\sum_{j\neq i}P_{ij}^2\frac{\Delta (Z_{i}\pi)^2 }{M_{ii}^2}\E [v_j\eta_j]\to^p 0.
\]
Finally, given that $\E [v_j\eta_j]$ is bounded by the constant we get that the last expression is bounded by $\Delta\frac{\pi'Z'Z\pi}{K}\to 0$. $\Box$

\begin{lemma}\label{lemma- properties of A}
Let $A=P^\perp-M_WD_\theta M_W$ where $\theta_1,...,\theta_N$ are selected in such a way that $A$ has all zero elements on the diagonal. Specifically $\theta=(M_W\circ M_W)^{-1}diag (P^\perp)$.
Then under Assumption~\ref{assumption- many controls} we have
\begin{itemize}
    \item[(i)] $c<\frac{1}{K_Z}\sum_{i,j}A_{ij}^2<C$; 
    \item[(ii)] $\frac{1}{K_Z^2}\sum_{i,j,k}A_{ij}^2A_{ik}^2\to 0$;
    \item[(iii)] $\frac{1}{K_Z^2}\sum_{i,j}A_{ij}^4\to 0$.
\end{itemize}
\end{lemma}

\paragraph{Proof of Lemma  \ref{lemma- properties of A}.}
First we notice that Assumption~\ref{assumption- many controls} implies that $\theta_i\leq \delta$. Indeed $\theta_i$'s are the solution to a system of linear equations $\sum_jM_{W,ij}^2\theta_j=P^\perp_{ii}$ and specifically $\theta=(M_W\circ M_W)^{-1}diag (P^\perp)$. Here matrix $M_W\circ M_W$ with elements $M_{W,ij}^2$ is diagonal-dominant as $\sum_{j\neq i}M_{W,ij}^2=M_{W,ii}-M_{W,ii}^2\leq M_{W,ii}^2$ under Assumption~\ref{assumption- many controls} and thus is invertible. Furthermore, the system of equations can be rewritten as $\theta_i=\frac{1}{M_{W,ii}^2}(P_{ii}^\perp-\sum_{j\neq i}\theta_jM_{W,ij}^2).$ Thus, if $\theta_i\geq 0$ for all $i$ as prescribed by Assumption~\ref{assumption- many controls} then $\theta_i\leq \frac{P_{ii}^\perp}{M_{W,ii}^2}\leq \delta$.

Below we use projection property $M_WP^\bot=P^\bot M_W=P^\bot$ and  $M_W^2=M_W$ as well as the definition of matrix $A$:
$ A_{ij}=P^\perp_{ij}-\sum_kM_{W,ik}\theta_kM_{W,kj}$. For part (i) notice that
\begin{align*}
    \sum_{i,j}A_{ij}^2=&\sum_i(P_{ij}^\perp)^2-2\sum_{i,j,k}P^\perp_{ij}M_{W,ik}\theta_kM_{W,kj}+\sum_{i,j,k,n}M_{W,ik}\theta_kM_{W,kj}M_{W,in}\theta_nM_{W,nj}=\\
    =&\sum_jP_{jj}^\perp-2\sum_{j,k}P^\perp_{jk}\theta_kM_{W,kj}+\sum_{j,k,n}M_{W,jk}M_{W,kn}M_{W,nj}\theta_k\theta_n\\
    =&K_Z-2\sum_k P_{kk}^\perp\theta_k+\sum_{k,n}M_{W,kn}^2\theta_k\theta_n\\= &K_Z-2\sum_k P^\perp_{kk}\theta_k+\sum_{k}P^\perp_{kk}\theta_k= K_Z-\sum_k P_{kk}^\perp\theta_k.
\end{align*}
Given $0\leq\theta_i\leq \delta$ we have $(1-\delta)K_Z\leq \sum_{i,j}A_{ij}^2\leq K_Z$.

For (ii) notice that:
\begin{align*}
\sum_iA_{ij}^2=&P_{jj}^\perp-2\sum_{k}P_{jk}^\perp\theta_kM_{W,kj}+\sum_{k,n}M_{W,jk}M_{W,kn}M_{W,nj}\theta_k\theta_n=\\
    =&P_{jj}^\perp-2\sum_{k}P_{jk}^\perp\theta_kM_{W,kj}+\sum_{k}M_{W,jk}\theta_k(P^\perp_{jk}-A_{jk});
\\
    \sum_{i,j,k}A_{ij}^2A_{ik}^2=&    \sum_i(P_{ii}^\perp-\sum_{k}P_{ik}^\perp\theta_kM_{W,ki}-\sum_{k}A_{ik}\theta_kM_{W,ki})^2.
\end{align*}
The sum above has terms: 
\begin{align*}
&\sum_i(P_{ii}^\perp)^2\leq P_{ii}^\perp\leq K_Z ;\\
    &\sum_{i}\left(\sum_kP_{ik}^\perp\theta_kM_{W,ki}\right)^2\leq \sum_i\left(\sum_{k}(P_{ik}^\perp)^2\sum_kM_{W,ki}^2\right)\max_k\theta_k^2\leq C\sum_iP_{ii}^\perp\leq CK_Z ;\\
    &\sum_{i,k}P_{ii}^\perp P_{ik}^\perp\theta_kM_{W,ki}\leq \sum_iP_{ii}^\perp\sqrt{\sum_k(P_{ik}^\perp)^2\sum_k M_{W,ki}^2}\max_k|\theta_k|\leq CK_Z;
\\
    &\sum_{i}\left(\sum_kA_{ik}\theta_kM_{W,ki}\right)^2\leq \sum_i\left(\sum_{k}A_{ik}^2\sum_kM_{W,ki}^2\right)\max_k\theta_k^2\leq C\sum_{i,k}A_{ik}^2\leq CK_Z;\\
       &\sum_{i,k}P_{ii}^\perp A_{ik}\theta_kM_{W,ki}\leq\sqrt{\sum_{i}(P_{ii}^\perp)^2}\sqrt{\sum_{i}\left(\sum_kA_{ik}\theta_kM_{W,ki}\right)^2}\leq CK_Z;\\
    &\sum_i\left(\sum_kP_{ik}^\perp\theta_kM_{W,ki}\right) \left(\sum_kA_{ik}\theta_kM_{ki}\right)\leq \sqrt{\sum_i\left(\sum_kP_{ik}^\perp\theta_kM_{W,ki}\right)^2} \sqrt{\sum_i\left(\sum_kA_{ik}\theta_kM_{W,ki}\right)^2}\leq CK_Z.
\end{align*}
Putting all terms together we get  $\sum_{i,j,k}A_{ij}^2A_{ik}^2\leq CK_Z$.

For (iii) notice that $$\sum_k|M_{W,ik}M_{W,jk}|\leq\sqrt{\sum_{k}M_{W,ik}^2\sum_kM_{W,jk}^2} \leq\sqrt{M_{W,ii}M_{W,jj}}\leq 1.$$ 
Thus
$|A_{ij}|=|P_{ij}^\perp-\sum_kM_{W,ik}M_{W,jk}\theta_k|\leq 1+\max_k|\theta_k|\leq C$.
This implies $$\sum_{i,j}A_{ij}^4\leq C^2\sum_{i,j}A_{ij}^2\leq CK_Z,$$ and (iii) holds. $\Box$

\paragraph{Proof of Theorem  \ref{thm- AR many controls}.} Statements proved in Lemma \ref{lemma- properties of A} alone with Assumption 2 lead to the validity of all conditions of the Central Limit Theorem for quadratic forms stated  in Corollary A2.8 in \cite{solvsten2020robust}. This implies that under the null ($H_0:\beta=\beta_0$) we have:
$$
\frac{1}{\sqrt{K_Z\Phi_W}}(Y-\beta_0 X)'A(Y-\beta_0X)=\frac{1}{\sqrt{K_Z\Phi_W}}e'Ae\Rightarrow N(0,1),
$$
when $N,K_Z\to\infty$ with
$
\Phi_W=\frac{2}{K}\sum_{i,j}A_{ij}^2\sigma_i^2\sigma_j^2,
$
and $\sigma_i=\E e_i^2$. What is left to prove is the consistency of $\widehat\Phi_W$, the estimator for $\Phi_W$ defined in equation (\ref{eq: def PhiW}). For this proof we follow closely the structure of the proof of Lemma 2 in \cite{mikusheva_inference_2022}. Specifically, define $M_{ij}=M_{ZW,ij}$, $\widetilde A_{ij}^2=\frac{A_{ij}^2}{M_{ii}M_{jj}+M_{ij}^2}$, and we want to show that
$$
\frac{2}{K}\sum_{i,j}\widetilde A_{ij}^2 e_iM_iee_jM_je-\frac{2}{K}\sum_{i,j}A_{ij}^2\E e_i^2\E e_j^2\to^p 0.
$$
For this we first notice that 
$$
\E\frac{2}{K}\sum_{i,j}\widetilde A_{ij}^2 e_iM_iee_jM_je=\frac{2}{K}\sum_{i,j}A_{ij}^2\E e_i^2\E e_j^2,
$$
define $\xi_{ij}=e_iM_iee_jM_je-\E[e_iM_iee_jM_je]$. Our goal is to show that $\frac{1}{K}\sum_{i,j}\widetilde A_{ij}^2\xi_{ij}\to^p 0. $ The covariance structure of $\xi_{ij}$ including statements that $\max_{i,j}\E\xi_{ij}^2 <C$ and $\max_{i,j,k}|\E \xi_{ij}\xi_{ik}|<C$ are proven in Lemma 2 of  \cite{mikusheva_inference_2022}. Following the proof of Lemma 2 of  \cite{mikusheva_inference_2022} the only conditions needed are $\sum_{i,j}A_{ij}^4\leq CK_Z$ and $\sum_{i,k,j}A_{ij}^2A_{ik}^2\leq CK_Z$, both of which are proven in Lemma \ref{lemma- properties of A}. $\Box$

\end{document}